\documentclass{article}

\usepackage{PRIMEarxiv}

\usepackage[utf8]{inputenc} 
\usepackage[T1]{fontenc}    
\usepackage{hyperref}       
\usepackage{url}            
\usepackage{booktabs}       
\usepackage{amsfonts}       
\usepackage{nicefrac}       
\usepackage{microtype}      
\usepackage{lipsum}
\usepackage{fancyhdr}       
\usepackage{graphicx}       
\graphicspath{{media/}}     

\title{Toward an auditory Virtual Observatory (preprint)\thanks{Submitted to J. Audio Eng. Soc., 2023 August. Published in JAES Vol. 72, No. 5, pp. 341-351, May 2024.

\url{https://doi.org/10.17743/jaes.2022.0146}.
Corresponding author: adrian.griber@alumnos.upm.es}}

\author{
  Adrián García Riber \\
  ETSIST\\
  Universidad Politécnica de Madrid \\
  28031, Spain\\
  \texttt{adrian.griber@alumnos.upm.es} \\
   \And
  Francisco Serradilla \\
  ETSISI\\
  Universidad Politécnica de Madrid \\
  28031, Spain\\
  \texttt{francisco.serradilla@upm.es} \\
}
  
\begin{document}
\maketitle

\begin{abstract}
The exploration of the universe is experiencing a huge development thanks to the success and possibilities of today's major space telescope missions which can generate measurements and images with a resolution 100 times higher than their precedents. This big ecosystem of observations, aimed at expanding the limits of known science, can be analyzed using personal computers thanks to the implementation of interoperable Virtual Observatory (VO) technologies, and massive portals of stellar catalogs and databases. In this context of global analysis of astronomical big data, sonification has the potential of adding a complementary dimension to visualization, enhancing the accessibility of the archives, and offering an alternative strategy to be used when overlapping issues and masking effects are found in purely graphical representations. This article presents a collection of sonification and musification prototypes that explore the case studies of the MILES and STELIB stellar libraries from the Spanish Virtual Observatory (SVO), and the Kepler Objects of Interest light curve database from the Space Telescope Science Institute archive (STScI). The work makes use of automation, machine learning, and deep learning algorithms to offer a ``palette" of resources that could be used in future developments oriented towards an auditory virtual observatory proposal. It includes a user study that provides qualitative and quantitative feedback from specialized and non-specialized users in the fields of Music and Astronomy.
\end{abstract}


\section{INTRODUCTION}
The current development of massive astronomical archives and virtual observatory technology (VO)~\cite{djorgovski2005virtual} offers a wide range of data products and services that can be explored through interoperable technology using a personal computer. Tools such as the \textit{ALADIN} sky Atlas~\cite{ref_Aladin, bonnarel2000aladin,boch2014aladin}, which provides direct access to worldwide active astronomical servers; \textit{Simbad}~\cite{ref_Simbad,wenger2000simbad}, which offers information about the astronomical objects studied in scientific articles; \textit{Vizier}~\cite{ref_Vizier}, that serves a complete library of published astronomical catalogs; \textit{TOPCAT}~\cite{ref_Topcat,taylor2005topcat}, that allows the edition and interactive graphical visualization of tabular data;  or the VO Sed Analyzer \textit{VOSA}~\cite{ref_Vosa,bayo2008vosa}, that calculates estimations of the mass and age of target sources, they all represent good examples of the resources that astronomers use in the analysis of their case studies.

Based on this environment of universal access, the use of sonification and musification in multimodal displays for the exploration of astronomical data offers an additional domain complementary to visualization, that allows researchers to get immersed in their case studies, identify patterns, trends and outliers in the datasets~\cite{hall2019design}, and navigate the massive downloads of big data generated by space telescopes~\cite{ref_Chandra}. It also makes stellar catalogs and databases more accessible for blind and visually impaired (BVI) users~\cite{noel2022accessibility}. 

The ecosystem of projects devoted to astronomical data sonification has been recently pictured by studies such as Zanella et al. (2022)~\cite{zanella2022sonification}, and initiatives such as the United Nations Office for Outer Space Affairs' Sonification report~\cite{ref_Unoosa}, materializing the interest of the international community in the use of sound as a way to contribute towards an inclusive and accessible future of Space Sciences and Astronomy~\cite{zanella2022sounds}.

Projects such as the Audible Universe~\cite{misdariis2022sound, harrison2022audible} or Astro Accesible~\cite{ref_Astroaccesible}, analysis tools such as xSonify~\cite{candey2006xsonify, diaz2011sonification}, Astronify~\cite{ref_Astronify} or Highcharts Sonification Studio~\cite{cantrell2021highcharts}, and collaboration networks such as the Sonification World Chat~\cite{ref_SWC}, are only a small sample of the efforts invested by international researchers for the acceptance and establishment of auditory representations in scientific contexts. 

Aligned with the multiple goals of Bardelli et al.~\cite{bardelli2021sonification} and De Campo's motivation to sonify scientific data~\cite{de2009science}, the creation of specialized tools for the auditory display of an object of investigation has also the potential to build natural interdisciplinary connections between the data and the sound properties, broadening the possibilities in the representation and perceptualization of scientific information~\cite{ivapp17}. 

In addition, considering ``sound as malleable material"~\cite{franinovic2013sonic}, the sonification of astronomical data opens new horizons for creativity applied to music and sound art, which are proven values in science communication and outreach~\cite{metallinou2022sonification}. 

This article explores the potential of multimodal astronomical data representations from a technology-driven perspective, providing tools adapted to each dataset, that could be used to enhance the accessibility of stellar libraries, to generate an additional dimension of the data through sound, and to convey interest in science. Potential final users include scientists interested in complementing the graphical representations of their case studies with sonification, scientific communicators, STEM educators, and professors aimed at engaging their students with Astronomy and Astrophysics through sound, and music creators or composers willing to bring science to the general public using astronomical data-driven musical compositions. The overall purpose of the work is to provide different strategies and designs for the conversion of stellar spectra and light curves into sound, for which two parallel lines of research have been developed. The first one is aimed at studying the possibilities of Sonification for the representation of astronomical data in scientific contexts. The second one is focused on the generation of astronomical data-driven musical compositions. Both research lines make use of machine learning algorithms to reduce the data, and to generate autonomous music scores from stellar libraries and catalogs. In Sec. 1, the two main types of data used in the research are introduced. Sec. 2 offers two designs for fast auditory exploration of astronomical databases based on parameter mapping. Sec. 3 is focused on the sonification of galaxy spectra.  Sec. 4 and Sec. 5 propose the use of deep learning and neural networks for the sonification and musification of stellar spectra catalogs. Sec. 6 provides an evaluation of the designs, and Sec. 7 concludes by summarizing the results. All the prototypes and data described in this article are publicly available at: \url{https://github.com/AuditoryVO}

\begin{figure}
\centering
\includegraphics[width=82mm]{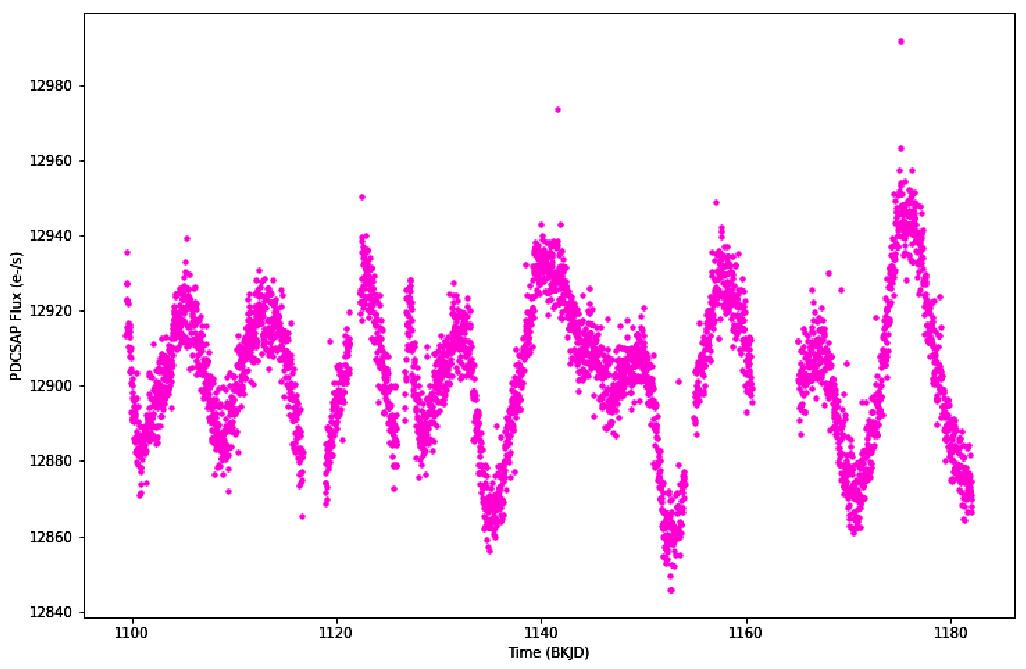}
\caption{Light curve of the rotating variable star KIC 11013096 with coordinates RA:18:47:44.69 and DEC:+48:30:29.63. Flux variations versus time. STScI.} 
\label{Lightcurve example}
\end{figure}

\section{DATA OVERVIEW}

\textit{Figures 1 and 2} show, respectively, an example of a stellar light curve and a spectrum. Light curves are two-dimensional graphical representations of the brightness flux variations over time, observed from an object of investigation. The analysis of these time series, mainly focused on the detection and evaluation of periodicity, is commonly used for the classification of eclipsing binaries, variable stars and supernovas, as well as for the discovery of extra-solar planets based on the transit detection method. Stellar spectra are graphical representations of the light flux variations of an object of investigation versus wavelength. Their analysis is used to determine the chemical composition, temperature, mass, and density of stars, galaxies, and exoplanets. Both are mainly stored and distributed in the standard \textit{Flexible Image Transport System (FITS)}~\cite{ref_FITS} file format.

\begin{figure}
\centering
\includegraphics[width=82mm]{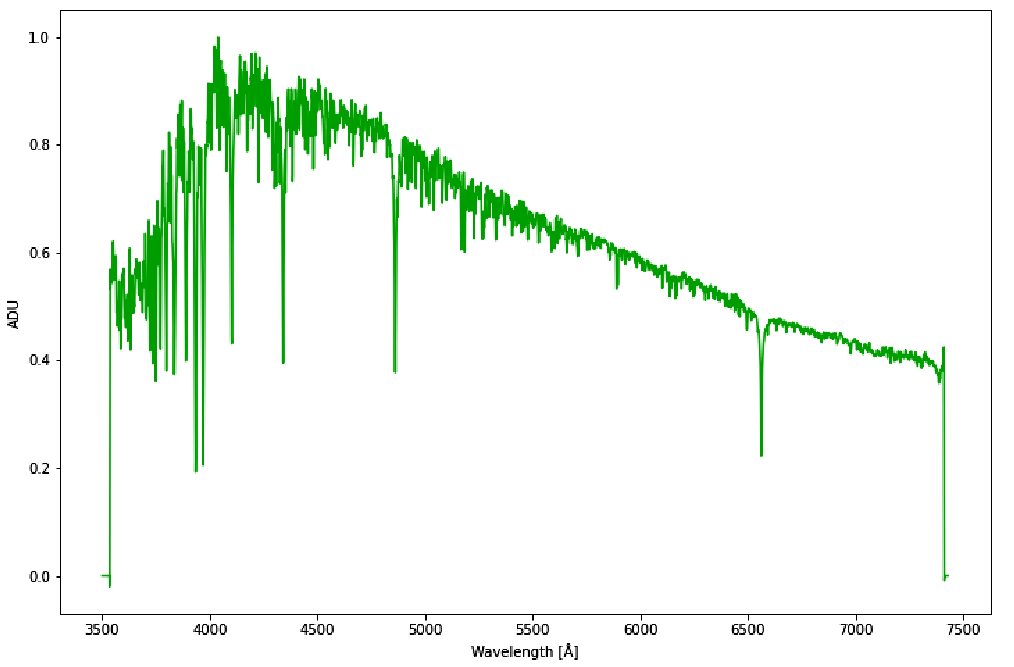}
\caption{Spectrum of the high proper motion star HD 125451 with coordinates RA:14:19:16.27 and DEC:+13:00:15.48. Flux variations versus wavelength. SVO.} 
\label{Spectrum example}
\end{figure}

\section{SEQUENTIAL AUDITORY EXPLORATION OF ASTRONOMICAL DATABASES}
Parameter mapping~\cite{grond2011parameter} is one of the most straightforward and commonly used sonification strategies, which consists of establishing a one-to-one correspondence between the properties of the data and the properties of the sound~\cite{barrass1999using}.  Analyzing the ``conceptual data dimension" to match listener expectancy~\cite{walker2002magnitude}, considering psycho-acoustics sensations generated by the sonification~\cite{ferguson2017evaluation}, and designing the polarity and scaling of the mapping are crucial for the  effectiveness and usefulness of a final auditory representation~\cite{axon2018sonification}. In the following sections we propose two different strategies for the sonification of RAW stellar spectrum and pre-processed light curve datasets.

\subsection{Mono-aural scanning}

Aimed at providing a first approach for the sequential sonification of lightcurves and stellar spectra, \textit{AutoFITS2Sound} converts the brightness flux variation of data streams into OSC messages~\cite{ref_OSC}, serializing the \textit{FITS2OSC}~\cite{garcia2022fits2osc} workflow. The pipeline allows sending and mapping accurate numeric values without note or scale constraints, and with a controlled precision degree that provides a perfect bridge for direct transduction between astronomical variables and sound parameters. Conceived as a tool for scanning stellar libraries, and based on the auditory graph philosophy~\cite{walker2010universal}, the prototype performs a simple x-to-time, y-to-frequency mapping that controls a single sinusoidal oscillator. It allows the representation of any kind of bi-dimensional data, such as light curves and stellar spectra.

The next video shows the \textit{AutoFITS2Sound} prototype during the exploration of 256 stellar spectra from the STELIB library~\cite{ref_STELIB} (SVO). As can be noticed, the absorption and emission lines in the spectra are easily identified as deep drops and peaks in pitch along the timeline. The representation also allows to understand the balance of energy in the spectrum as function of wavelength, something that could be useful for auditory spectra classification.

\url{https://github.com/AuditoryVO/AutoFITS2Sound}

\url{https://vimeo.com/641208770}

\subsection{Multi-variable soundscapes}
In a completely different approach, the \textit{DVT Explorer} prototype, fully described in the article \textit{Sonification of TESS data validation time series files}~\cite{garcia2022sonification}, generates sequential representations of interactive complex soundscapes. This design uses star and planet variables extracted from the \textit{Data Validation Time-series }(DVT) files of the \textit{Transiting Exoplanet Survey Satellite} (TESS) mission. It provides multi-variable auditory representations of the effective temperature, metallicity, surface gravity and stellar radius of host stars, as well as the orbital period, transit duration and depth of planet transits included in the TESS Object of Interest (TOI) catalog. 

Two versions of the prototype have been implemented. The first one is aimed at providing a sequential exploration of the catalog. As can be seen in the first demonstration video provided below, the duration and depth of the transits drive, respectively, the duration and amplitude of the sounds, offering a multimodal representation that has certain reminiscences with the \textit{Spectralism}~\cite{harvey2000spectralism}. The second design allows the analysis of single objects using VO tools and queries. 

\url{https://github.com/AuditoryVO/DVT-Explorer}

\url{https://vimeo.com/728364932}

\url{https://vimeo.com/723815913}

\section{GALAXY SPECTRA SONIFICATION}
The absorption and emission lines drawn in a spectrum represent the wavelengths at which chemical elements absorb or emit light. Their analysis is not only used in the study of star characterization and classification, but also applied to complete galaxy spectra as a way to get ``a powerful diagnostic of its stellar content and evolutionary properties", Kennicutt (1991)~\cite{kennicutt1992spectrophotometric, kennicutt1992integrated}. Promising results on galaxy spectra sonification have been achieved with spectral audification~\cite{trayford2023inspecting}, applying a direct conversion of data into sound to explore datacubes.

In a completely different approach that makes use of pre-processed data, the \textit{AbsEmis2Sound} prototype  is proposed for the sonification of Large Early Galaxy Census (LEGA-C) spectra~\cite{ref_Lega}. This prototype is oriented to provide a simple sonification of the absorption and emission lines of the spectra, that can reveal the existence of additional hidden lines overlapped in their graphical representations. This strategy could be useful in scientific data analysis, and could provide a way of making this analysis more accessible for BVI researchers.

Attending the mappings used in this prototype, a direct wavelength-to-frequency strategy was adopted to make the model as intuitive as possible. The spectroscopic analysis include smoothing, continuum fitting and substraction for the final line identification is described as part of STScI's post-pipeline Data Analysis Tools Ecosystem~\cite{ref_spectroscopy}. The first video provided bellow shows how the design allows final users to switch between the representation of pre-calculated emission and absorption lines, to analyze their number, amplitude, and frequency. In the second video, four different approaches are presented. The first representation generates a simultaneous chord with all the emission/absorption lines, in which each line is represented by a different musical instrument. The second provides a similar representation using sine waves. The "beating" effect produced when two or more sine waves are close in frequency helps in the detection of additional emission/absorption lines, but to accurately perceive their number, another kind of approach is needed. One solution is to present the lines sequentially. The third example shows this kind of sonification using piano notes, with the pitch of each note corresponding directly to the wavelength of each absorption/emission line. This approach makes it possible to count the events, but those lines that are close in frequency appear represented with the same note due to MIDI quantization. The final example performs a sequential representation with sine waves, which improves close frequency event differentiation and amplitude appreciation. The sound events have been also panned from left to right to clarify their identification.  

\url{https://github.com/AuditoryVO/AbsEmis2Sound}

\url{https://vimeo.com/771366086}

\url{https://vimeo.com/793542919}

\section{DEEP LEARNING SONIFICATION}

The use of deep learning techniques in automatic feature extraction from astronomical data is currently generating interesting results on the estimation of atmospheric parameters from stellar spectra~\cite{yang2015autoencoder}. The application of reduction techniques based on autoencoder architectures to large astronomical datasets also has the potential to improve the exploration and classification of the inherent complexity of galaxies~\cite{portillo2020dimensionality}, allowing the capture of intrinsic spectral features regardless of redshifts (stretch of the spectrum due to the wavelength increase generated by the increasing distance to the observed object), noise, and artifacts present in the data~\cite{melchior2022autoencoding}. 

\begin{figure}
\centering
\includegraphics[width=130mm]{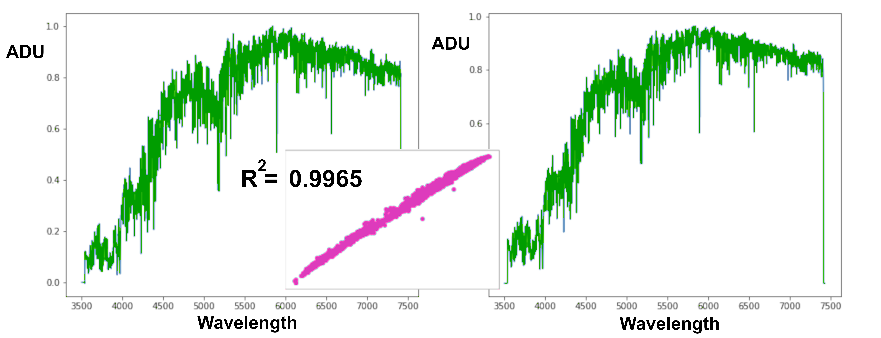}
\caption{Graphical comparison of the results of a four-layer four-dimensional variational autoencoder. Input spectrum (left) and the decoded output (right) for the \textit{K} type high proper motion star HD184406, with coordinates RA:19:34:05.35 and DEC:+07:22:44.18. R\textsuperscript{2} deviation = 0.9965.}
\label{Autoencoder}
\end{figure}

Autoencoders are formed by an encoder and a decoder. Both are neural networks with a variable number of hidden layers, trained to output an imperfect copy of the input by reducing its dimension to a compressed representation called \textit{latent space}~\cite{goodfellow2016deep}.

Variational autoencoders are a subcategory of autoencoders in which the latent vectors obtained are constrained to a continuous probability distribution, making it possible to generate realistic synthetic outputs through the exploration of their latent vectors. This allows the generation of accurate synthetic light curves with desirable characteristics used to characterize distant exoplanets and stars~\cite{woodward2019generating}, and to augment neural network training sets.

\subsection{Stellar spectra additive synthesizer}

\textit{Autoencoder2Sound} is a software prototype that synthesizes musical notes and chords from the sonification of the ten-dimensional latent space of a sparse autoencoder. Aimed at providing a method to represent with accuracy all the information contained in a stellar spectrum, the model is able to reduce the 4367 flux values of each spectrum to a vector of 10 dimensions. These vectors are sonified and presented sequentially in an additive audio synthesis philosophy. As can be seen in \textit{Figure 3}, the results on feature extraction in stellar spectra are promising despite the simplicity of the network, and the small number of curves used for the training. Alternative corpus and network configurations have been tested, finding that the model is highly dependent on the input data.

Attending the mapings and technical details of the design, the model has been trained with 985 stellar spectra from the MILES stellar library~\cite{ref_Miles, sanchez2006medium} (SVO). The encoder, which generates the latent space trying to mimic each stellar spectrum, uses two intermediate \textit{Dense} layers and trains 2,099,350 parameters during 100 epochs. The neural network has been implemented using \textit{TensorFlow2}~\cite{ref_Tensor, tensorflow2015-whitepaper}, and the event generation is done in \textit{Python} and sent to \textit{Cabbage-CSound} via \textit{OSC}. The \textit{Loss} function error obtained was 0.0022. 

From the audio generation perspective, this prototype can be seen as a 10 sinusoids additive synthesizer. Each dimension of the autoencoder has been directly mapped to the frequency of each oscillator, multiplied by a factor of 10,000 to bring it to the audible range. All the amplitudes of the sound generated are equal, and the resulting fundamental frequencies below 20Hz have been filtered. The image presented on the left is the original spectrum. The image on the right is the image recovered from the sonified latent vector that is provided for comparative purposes.

Two versions of the synthesizer have been developed. The first one, \textit{Autoencoder2Chords}, presents the 10 oscillators sounding simultaneously in a ``chordified" representation. The second one, \textit{Autoencoer2Notes}, activates sequentially each oscillator to offer a complete flux of notes between spectra.  The prototypes and demonstration videos of both versions are available at:

\url{https://github.com/AuditoryVO/Autoencoder2Sound}

\url{https://vimeo.com/641219356}

\url{https://vimeo.com/641218256} 

\subsection{Stellar spectra latent space spatialization}
\textit{VAE2Sound} is a prototype that explores the concept of stellar spectra-driven latent space sonification using \textit{variational autoencoders}. It provides a method for reducing the data of the spectra in a similar way to the previously described design. Nevertheless, the use of a variational autoencoder allows the generation of synthetic stellar spectra by sampling its latent space, which could be useful in data augmentation. The main goal of this design is to provide a spatialized surround representation of the spectra that could be used in public presentations, conferences or planetarium exhibitions.

In this design, the 12 values of each autoencoder's latent space vectors are mapped to the fundamental frequencies of 12 independent sine generators. Each of these oscillators is sent to a loudspeaker to allow the acoustic spatialization of the representation in a surround sound system formed by 12 loudspeakers. One interesting thing about this spatialization is that it can also be seen as a spatial representation of the \textit{latent space}, providing a way of monitoring the behavior of the network. The model has been trained during 100 epochs with the MILES library for comparative purposes. The loss function value obtained is 0.0158, finally reduced to 0.0043 after an optimization of the learning rate. The mapping strategy used is identical to the previously described approach, converting the latent values to scaled fundamental frequencies for 12 independent oscillators. The latent values were multiplied by 10,000, and the frequencies below 20Hz were also filtered.

The graphical display of the prototype provides the original spectrum, the recovered spectrum from the latent vector which is sonified, and a representation of the active speakers. Mute buttons are incorporated on each channel to allow a controlled monitorization. The prototype, and a demonstration video showing the stereo bounced exploration of 25 stellar spectra from the MILES library are available at: 

\url{https://github.com/AuditoryVO/VAE2Sound}

\url{https://vimeo.com/641221857}

 

\begin{figure}
\centering
\includegraphics[width=90mm]{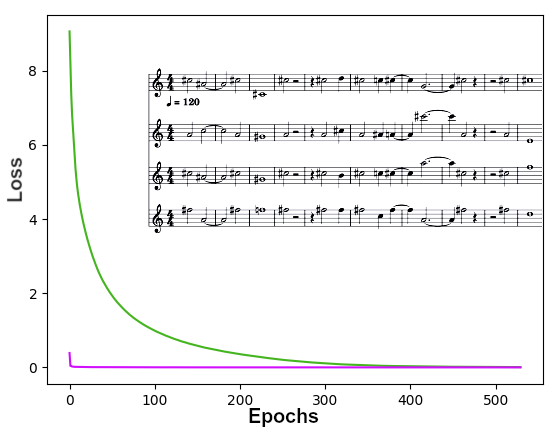}
\caption{Results of the LSTM with attention generative model used for stellar spectra-driven unsupervised music composition. Evolution of pitch loss (green curve) and note duration loss (fuchsia curve) during 525 training epochs (four and a half hours of calculations using a high performance 24GB GPU).}
\label{Learning_curve}
\end{figure}
 
\section{STELLAR SPECTRA-DRIVEN UNSUPERVISED MUSIC COMPOSITION}
To conclude this approach to the interdisciplinary possibilities involving stellar catalog exploration, deep learning techniques, and music generation, an additional research line aimed at generating unsupervised original music pieces from stellar spectra has been developed. All the details of the automatic composition system designed are published in the article \textit{AI-rmonies of the Spheres}~\cite{riber2023ai}. This approach uses a Long Short Term Memory (LSTM) with attention neural network architecture to generate a not sounding score in the style of a user-selected MIDI file. This LSTM output is cross-matched with a \textit{Pitch Class Set Theory} algorithm, that looks for ``matching chords" with those obtained by the ``chordification" of the MILES stellar library by a variational autoencoder. A final MIDI score with the ``matched stellar chords" is generated to be rendered by any user-selected software instrument or synthesizer. This approach allows the generation of completely original pieces, using the musical structure of the input as an underlying trigger that maintains musical consistency. The generated pieces show characteristic sound properties that do not correspond to the style of the piece of music used as input. The next video provides an example of an original unsupervised stellar spectra-driven composition generated using a MIDI trigger of J. S. Bach's \textit{Cello Suite No. 1 in G major}. The visuals synchronized with the music show a representation of each object and its spectrum, which actually drives the composition. After obtaining the unsupervised composition, a human process done by the authors has been introduced to select the instrumentation and final sound design of the piece.

\url{https://vimeo.com/835551607}

Regarding the technical details of the generative process, the 21,492,526 parameter LSTM with attention network used was trained over 1277 scores (82231 unique notes with 118 duration values), extracted from the MAESTRO corpus~\cite{hawthorne2018enabling}. The final \textit{Loss} function error was 0.0088  after 525 epochs using a learning rate of 0.001. \textit{Figure 4} shows the training curve performed by the model. The prototype can be downloaded from:

\url{https://github.com/AuditoryVO/AI-rmonizer}

\section{EVALUATION}

Aimed at evaluating the sonification prototypes described in previous sections, we provide a quantitative R\textsuperscript{2} analysis of the deep learning models used in section 4, as well as a survey with 25 questions to be answered by both, specialized and non-specialized users in the fields of Music and Astronomy.

\subsection{R\textsuperscript{2} deep learning sonification analysis}
The analysis of the coefficient of determination R\textsuperscript{2}~\cite{cameron1997r} between the original and the decoded spectra, provides a value between 0 and 1 that indicates the correlation of the original fluxes with those recovered from the calculated latent space, with 1 indicating that the decodification perfectly fits the input data, and 0 indicating an absolut mismatch. This coefficient has been calculated for the entire dataset, and for each decoded spectrum, to obtain a quantitative measure of the models' accuracy.

\begin{figure}
\centering
\includegraphics[width=100mm]{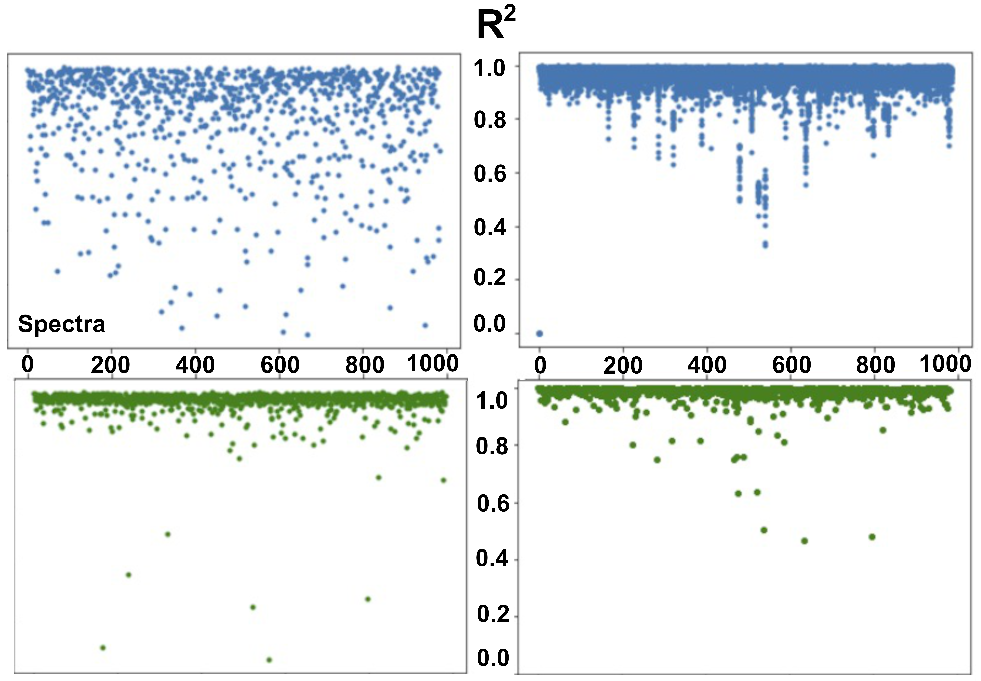}
\caption{Distribution of R\textsuperscript{2} results for VAE (blue/up), and sparse autoencoder (green/down), trained on the original dataset (left), and on the x20 augmented dataset (right). Spectrum label vs R\textsuperscript{2}.}
\label{R square}
\end{figure}

Training the VAE on the original MILES library that contains 985 spectra, we obtained a total R\textsuperscript{2} of 0.834 with 45,79 per cent of the spectra presenting a R\textsuperscript{2} higher than 0.9. Augmenting the dataset by repeating each spectrum 20 times, we obtained a total R\textsuperscript{2} of 0.976 for the complete dataset, with 96,92 per cent of the spectra presenting a R\textsuperscript{2} higher than 0.9. Minimum R\textsuperscript{2} = 0.3283,  maximum R\textsuperscript{2} = 0.9981, median = 0.9784, standard deviation = 0.038. 

Applying the same analysis to the sparse autoencoder using the same dataset, we obtained a total R\textsuperscript{2} of 0.96 with 94,11 per cent of the spectra presenting a R\textsuperscript{2} higher than 0.9 for the original dataset, and a total R\textsuperscript{2} of 0.985, with 97,868 per cent of the spectra presenting a R\textsuperscript{2} higher than 0.9 for the augmented dataset. Minimum R\textsuperscript{2} = 0.4648,  maximum R\textsuperscript{2} = 0.9983, median = 0.9743, standard deviation = 0.064.

\textit{Figure5} provides a comparison between the distribution of R\textsuperscript{2} results for the original and augmented datasets using variational and sparse autoencoders. Notice how the
sparse autoencoder provides better results for the original dataset. The computation time is also 10 times shorter for the sparse autoencoder (8 vs 83 s/epoch for the augmented dataset).

\subsection{Survey design}

To record quantitative and qualitative feedback of the designs described in this article, a survey link was distributed to potential volunteer participants among personal and professional contacts of the authors from 7 December 2022 to 25 January 2023. Participants were advised to use headphones, with no response time limit imposed. The questionnaire included two categories of questions using embedded videos to support the multimodal representations. It is available for reference at:

\url{https://forms.office.com/e/iCFyXfwsdT}

The first category was formed by 8 quantitative questions aimed at identifying the most effective sonification strategy for revealing the number of overlapped emission and absorption lines in galaxy spectra, using the Large Early Galaxy Census (LEGA-C) survey. The spectra were represented using the \textit{AbsEmis2Sound} prototype described in \textit{section 3}. Two additional questions oriented to analyze the ability of the participants in the recognition of absorption lines with the \textit{FITS2OSC} prototype described in \textit{section 2.1} were asked, using stellar spectra from the STELIB library.

The second category of questions was aimed at obtaining qualitative feedback on the different sonification and musification strategies provided, as well as on the stellar spectra-driven musical compositions proposed in \textit{section 5}. With the intention of opening the scope of the survey to less experienced participants, no special criteria were established for these questions, trying to capture first impressions of the participants, which could be useful in the design of widely accepted intuitive auditory representations. Additional categorical questions related to visual and auditory impairments, as well as experience level in the fields of Music and Astronomy, were added. All the questions were presented shuffled. 

\subsection{General results}

We received a total of 46 survey responses from participants with ages ranging between 18 and 65 years old, most of them residing in Spain, with some exceptions from Australia and USA. The answers were classified into four groups according to the experience of the participants in Astronomy and Music, understood as their familiarity with the nature of the astronomical data or their ability to pay attention to sound events. A total of 17 participants were classified as experienced, including 4 professional or amateur astronomers, 10 professional or amateur musicians and 3 people belonging to both groups who were considered as potential ``experts" on the task. The other 29 participants were included in the ``non experts" group. None of the participants declared hearing impairments. The participants were also divided into non-BVI and BVI groups, the latter including 17 people with slight vision loss and two blind participants. Although these numbers are unfortunately too low to make any statistically significant conclusion, the observed tendencies presented below could be useful as a preliminary guide for making design decisions in future developments. The complete dataset and the following analysis is available at: \url{https://github.com/AuditoryVO/Evaluation}

\subsection{Auditory recognition of absorption and emission lines}

The ten questions related to the auditory identification of spectral lines offered an average success rate of 0.26087 with a \textit{Jeffreys'} confidence interval of (0.24093, 0.28184), and a 68.3 percent of uncertainty, suggesting that the task was difficult for all participants.

\begin{figure}
\centering
\includegraphics[width=120mm]{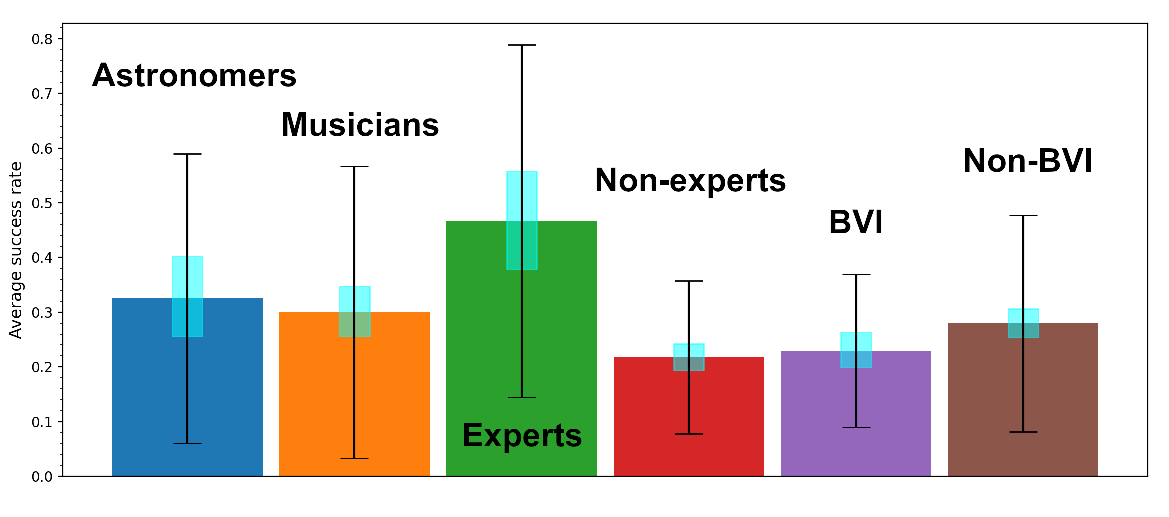}
\caption{Results for the auditory recognition of absorption and emission lines. Standard deviation represented in black and Jeffreys' confidence intervals in blue. Notice that the groups are not balanced to establish a direct comparison between them.}
\label{Quantitative results}
\end{figure}

Although in this preprint version of the article the groups are not balanced to establish direct comparisons, success rates obtained by experienced users in relation to non-experienced users are consistent with the conclusions reported by Walker and Nees (2011)~\cite{walker2011theory}, and Tucker(2022)~\cite{tucker2022evaluating}, suggesting that training and knowledge in the field of study are required to have a useful integration of Sonification in scientific research. Our initial thesis was that even non-expert participants would have been able to count sound events, but the results were not aligned with this idea, suggesting that even counting simple sound events requires training. As can be seen in \textit{Figure 6}, experienced participants in both fields, Music and Astronomy, obtained 2.14 times better results than non-experienced participants. The astronomers performed eight percent better than the musicians, and 1.5 times better than non-experienced participants. The extremely low success rates obtained in question 6 suggest that sequential approaches could be much more useful than ``chordified" approaches on the auditory recognition of multiple absorption and emission lines of galaxy spectra. The highest success rates were obtained in question 7, pointing to the sequential sine waves representation as the most useful and efficient sonification strategy for this task, where graphical representation overlays the lines. Non-BVI participants performed 21 percent better than BVI.

\subsection{Qualitative results}

A synthesis of the qualitative feedback recorded in the survey is presented in \textit{Figure 7}. The answers to the ten questions about development strategies and unsupervised music compositions have also been classified into four groups of participants according to their level of expertise. It is worth mentioning that the answers of the astronomers population clearly presented softer polarization than the rest of the groups, which could be related to aesthetics expectancy from musicians and non-experienced participants. In question number 10, which provides an evaluation of the autoencoder prototype on stellar spectra classification tasks, musicians and astronomers found the sine wave sonification more interesting than the \textit{chordified} piano representation of the M-K sequence. In the case of the astronomers it could suggest a clear preference for abstract scientific representations, while for musicians, it could be associated to the expectations associated to the instrument. This topic represents an additional line to be analyzed in future studies. Assuming the need of training that sonification and musification involve, we found a reference point on western contemporary music as a precedent with several decades of history, that required similar training for listeners, and that could be useful to understand what training for sonification means. 

A general tendency for all participants to give higher ratings to classic musical forms over abstract representations is found in the rest of the questionnaire, and can be appreciated with the comparison of the results from questions 18, 19, and 20. This tendency can also be appreciated in the subjective evaluation of the unsupervised music compositions of questions 21 and 22. The average result for all qualitative questions is 6.41 and the median is 7.0, which might suggest a good acceptance of the proposals.

\begin{figure*}
\centering
\includegraphics[width=\linewidth]{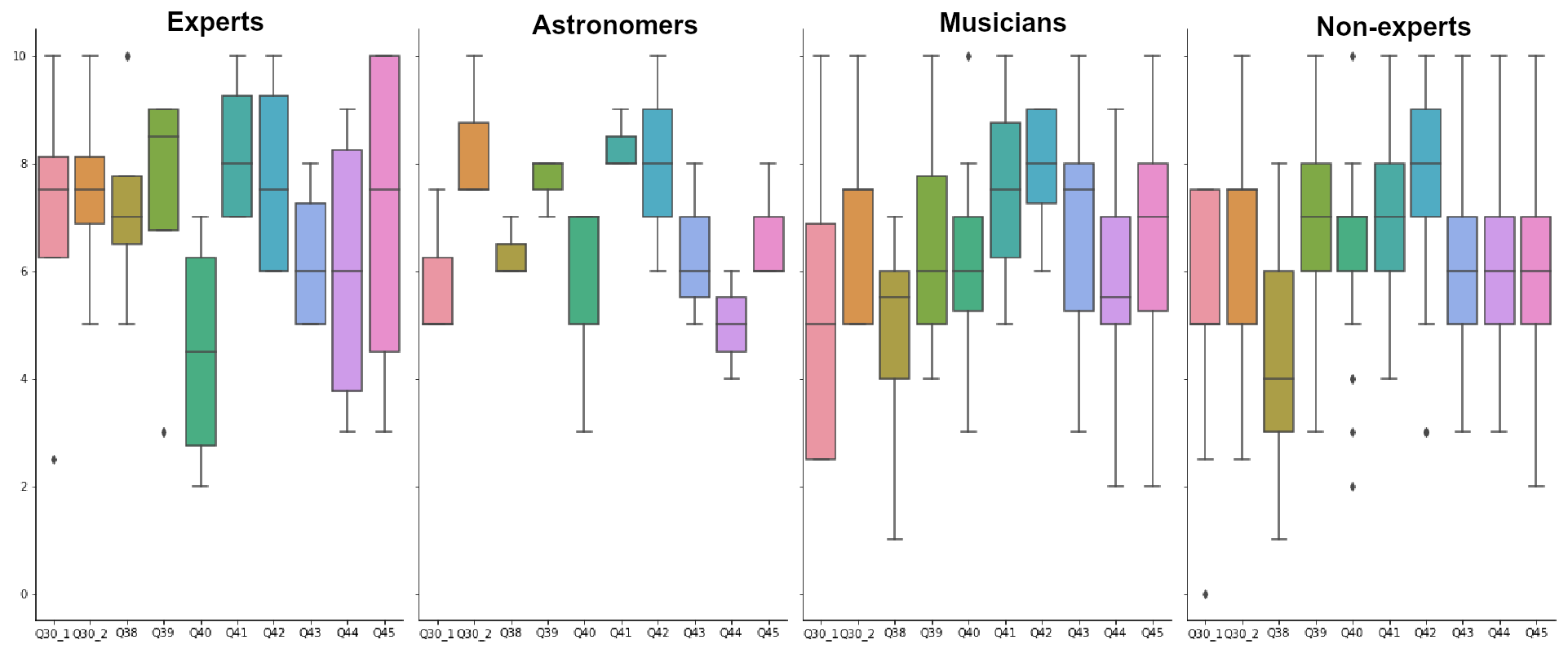}
\caption{Qualitative analysis by groups of experience in Music and Astronomy. Peak values, outliers, average, and error bars for the recorded responses. The ``experts" label corresponds to professional and amateur astronomers-musicians, the ``astronomers" group corresponds to professional and amateur astronomers, the ``musicians" category corresponds to professional and amateur musicians, and the ``non-experts" to those who do not fall into any of the previous categories. The questions referenced in the x-axis are available at: \url{https://forms.office.com/e/iCFyXfwsdT}} 
\label{Qualitative}
\end{figure*}

\section{CONCLUSION AND PROSPECTIVE}
The sonification and musification prototypes presented in this article offer a wide list of perspectives and open source resources that could be applied to different study techniques such as transit photometry, transmission spectroscopy, and spectrophotometry, which are respectively used in the search for planet transits, exoplanet atmosphere characterization, and star or galaxy formation, evolution, and classification. In addition to their technical description, a quantitative and qualitative evaluation conducted by specialized and non-specialized participants is provided, offering tendencies that could be useful as preliminary guidelines for future developments. These tendencies, summarized in the following paragraphs, should be considered indicative, since the number of participants did not provide statistical significance.

A deeper evaluation by a bigger group of experts is part of our current work, which is focused on the design of a user study with more scientific participants to analyze the inclusion of deep learning sonification into a scientific graphical display prototype for the exploration of the CALIFA survey dataset~\cite{garcia2014califa}. It is expected to confirm the promising results advanced in this study.

Incorporating auditory representations into scientific problems such as the analysis of absorption and emission lines of galaxy spectra, could provide useful complementary information to graphical representations, what could help in the detection of hidden events and overlapped information, and could make this analysis more accessible for BVI users.

The experience and knowledge of the users in both fields, Music and Astronomy, can contribute in the acceptance of sonification as an effective tool for the communication of astronomical information. Nevertheless, simple actions like counting sound events require training and concentration. The need for training that sonification and musification involve, could  benefit from the strategies adopted by western contemporary music as a precedent with several decades of history that required equivalent training for listeners. 

In this work we propose the use of sine waves for the representation of accurate scientific information, offering harmonic and enharmonic results, that acquire musical characteristics when presented sequentially or ``chordified". Simple aesthetics add-dons like \textit{Reverb} are used to emphasize this musicality, which takes this kind of representation close to electroacoustic music, reducing fatigue and enhancing its acceptance without any accuracy loss.

Alternatively, the MIDI protocol is also used to allow the musification and mapping of data to virtual musical instruments, and to autonomously generate music sheets. This method can be useful for science outreach and for the general public engagement with science in live commented concerts and conferences. Nevertheless, its intrinsic frequency quantification into musical notes, increases the uncertainty of the results, making it less suitable for accurate scientific applications. Some exceptions could include the use of auditory thumbnails as a potential first approach to massive archive data exploration, and the use of musical instrument mappings for timbre detection of near-coincident frequency events, such as the case study of the absorption and emission lines from galaxy spectra.

The unsupervised compositions and catalog musifications that make use of acoustic instruments and classical music forms obtained regularly higher scores in the qualitative evaluation, suggesting that abstract sound representations could be less suitable for outreach applications and for the general public.

As a general conclusion, further development and evaluation are needed to confirm the promising tendencies presented in this article, and to advance towards the improvement of the acceptance of auditory displays for scientific applications. However, the case studies and auditory designs here presented, which range from single sine wave mapping to the complete generation of unsupervised polyphonic musical compositions from stellar spectra, provide a broad ecosystem of tools that can be used in Music, Sound Design, Sound Art, Science Outreach, and STEM engagement activities. Additionally, these strategies could also be useful as reference points for future developments in the context of an astronomical auditory virtual observatory, that could allow the exploration of public stellar libraries and catalogs through sound, enhancing their accessibility, and improving inclusion in space science.

\section{ACKNOWLEDGMENT}

This work is based on data from the MILES and STELIB library service developed by the Spanish Virtual Observatory in the framework of the IAU Commission G5 Working Group: Spectral Stellar Libraries.

This research also includes data collected by the Kepler and TESS missions, obtained from the MAST data archive at the Space Telescope Science Institute (STScI). Funding for the Kepler mission is provided by the NASA Science Mission Directorate. Funding for the TESS mission is provided by the NASA Explorer Program. STScI is operated by the Association of Universities for Research in Astronomy, Inc., under NASA contract NAS 5-26555

\bibliographystyle{unsrt}
\bibliography{SonificationJAES}

\end{document}